\documentclass[aps,pre,twocolumn,amsmath,amssymb,showpacs,superscriptaddress,floatfix]{revtex4-1}
\usepackage{color}
\usepackage{graphicx,epsfig}
\usepackage{srctex}
\usepackage[hypertex]{hyperref}
\DeclareMathOperator{\Tr}{Tr}
\DeclareMathOperator{\RS}{\mathrm{\scriptscriptstyle RS}}\DeclareMathOperator{\1RSB}{\mathrm{\scriptscriptstyle 1RSB}}
\DeclareMathOperator{\RSB}{\mathrm{\scriptscriptstyle 1RSB}}
\DeclareMathOperator{\FRSB}{\mathrm{\scriptscriptstyle FRSB}}

\begin{document}

\title{Generalized Sherrington--Kirkpatrick-glass  without reflection symmetry }

\author{T.~I.~Schelkacheva}
\affiliation{Institute for High Pressure Physics, Russian Academy of Sciences, 142190, Troitsk, Moscow,
 Russia}
\author{E.~E.~Tareyeva}
\affiliation{Institute for High Pressure Physics, Russian Academy of Sciences, 142190, Troitsk, Moscow,
 Russia}
\author{N.~M.~Chtchelkatchev}
\affiliation{Institute for High Pressure Physics, Russian Academy of Sciences, 142190, Troitsk, Moscow,
 Russia}
\affiliation{L.D. Landau Institute for Theoretical Physics, Russian Academy of Sciences,117940 Moscow, Russia}
\affiliation{Department of Theoretical Physics, Moscow Institute of Physics and Technology, 141700 Moscow, Russia}
\affiliation{Department of Physics and Astronomy, California State University Northridge, Northridge, CA 91330, USA}

\begin{abstract}
We investigate generalized Sherrington--Kirkpatrick glassy systems without reflection symmetry. In the neighbourhood of the transition temperature we in general uncover the structure of the glass state building the full-replica-symmetry breaking solution.  Physical example of explicitly constructed solution is given.
\end{abstract}


\maketitle


\section{Introduction\label{Sec:Intro}}

Sherrington-Kirkpatrick (SK) model --- Ising model with random exchange interactions --  has proved to be unique laboratory for understanding physics of spin glasses~\cite{sk1975}. Its exact mean-field solution has become the cone-stone of modern glass-physics. There are many generalizations of SK-model that allow understanding other glassy systems, more complicated than spin glasses~\cite{gillin2001multispin,GillinSherringtonJphys2001Comment,Parisi2003,Montanari2003,Crisanti2005,Crisanti2006,Crisanti2007,zamponi2010,tareyeva2010spin,crisanti2010sherrington,Tareyeva2013PLA,VasinTMF2013,wolynes2012structural,2013arXiv1309.4292,Janis2013,Parisi2012PRB}.
One of the simplest ways to build the generalized SK model (GSK) is to replace Ising-spin operator at each lattice site, $\hat S_z$, by another diagonal operator $\hat U$ that also, of course, satisfies the relation, $\Tr \hat U=0$, as well as Ising-spin does. Then one may naively suggest that this new glass forming model inherits at least on qualitative level most of the features of the SK-model. However this is not so~\cite{lutchinskaia1984,GoldbartSherringtonJPhys1985,lutchinskaia1995,Schelkacheva2007PRB,Schelkacheva2008TMF2,SchelkachevaPRE2009,SchelkachevaPRB2010,Tareyeva2013PLA}. It follows that physics of generalized SK-model strongly depends on the hidden symmetry of $\hat U$-operator, particularly if there is ``reflection'' symmetry or not. Here we  build analytical solution for the glass state in the generalized SK-models where $\hat U$-operators do not have reflection symmetry.

Formally the presence of the reflection symmetry means that $\Tr[{\hat{U}}^{(2k+1)}]=0$, $k=0,1,2,\ldots$. Ising spin operator, $\hat S_z$, obviously satisfies the reflection symmetry condition. It was shown recently ~\cite{schelkachevaPLA2006} that all GSK-models with reflection symmetry qualitatively behave as the SK-model.

\begin{figure}[bh]
  \centering
  \includegraphics[width=0.9\columnwidth]{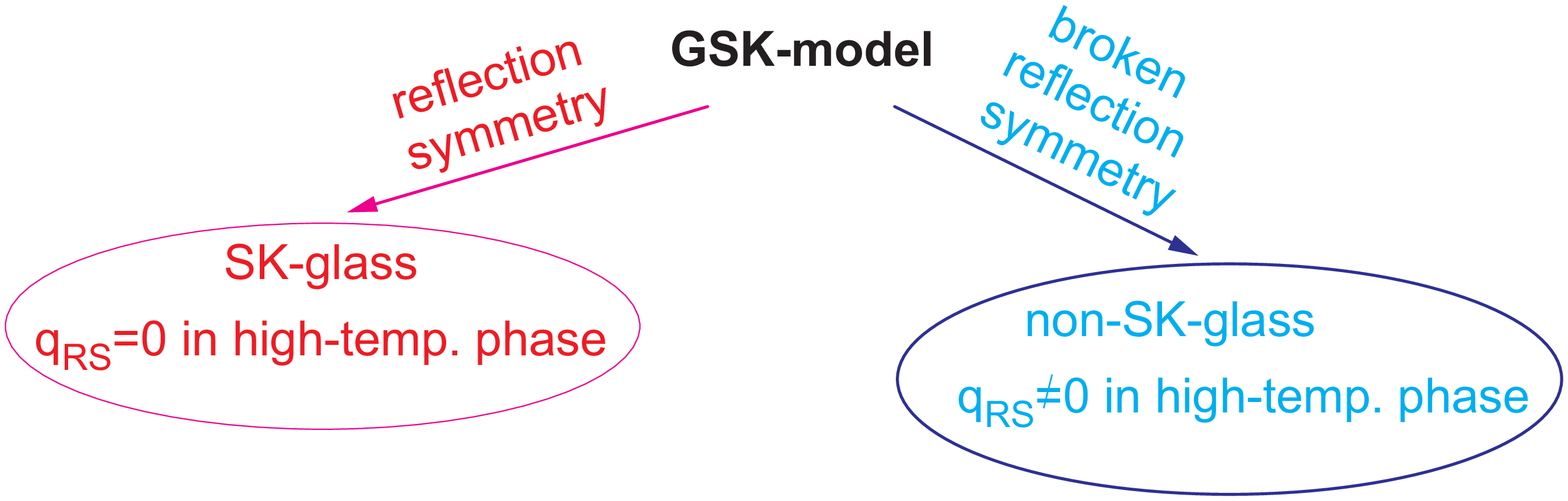}\\
  \caption{(Color online) GSK models with and without reflection symmetry produce principally different glass states. While symmetric GSK-glass is well understood, the nonsymmetric one --- does not~\cite{Tareyeva2013PLA}. One of the drastic differences between the two classes of GSK models is the behaviour of replica-symmetric \textit{glass} order parameter $q_{\RS}$ in para-state (at temperatures above the glass transition). Here we uncover the nature of nonsymmtric GSK-glass using the replica symmetry breaking language. }\label{fig0}
\end{figure}

$\Tr[{\hat{U}}^{(2k+1)}]\neq 0$, $k=1,2,\ldots$ in the GSK-model without the reflection symmetry. Typical example is, e.g., the quadrupole-SK, where $\hat U$ is the quadrupole angular moment operator. More examples can be found in Ref.~\onlinecite{Tareyeva2013PLA}.

The nonsymmetric GSK-model has different glass-structure than the reflection symmetric one, see Fig.~\ref{fig0}. Only recently the full-replica-symmetry breaking solution of the nonsymmetric GSK-glass has been build, but for very special case when $\hat U=\hat U_0+\hat U_{1}$, where $\hat U_{1}$ is much smaller (in some sense) than $\hat U_0$; here $\hat U_0$ is reflection symmetric diagonal operator while $\hat U_{1}$ is non-reflection-symmetric (diagonal) perturbation, see Letter~\cite{Tareyeva2013PLA}. In the present paper we in general uncover the structure of the glass state in the  non-reflection-symmetric problem without any simplifying assumption about the smallness of the nonsymmetric part of $\hat U$. Our investigation is focused on the region near the glass-transition temperature where we follow how the glass freezes and unfreezes when we cross the glass transition temperature.

The concept  glass transition as a replica symmetry breaking (RSB) has proven to be very successful~\cite{Parisi1979,mezard1987spin,parisi2002arxiv,gotze2008complex,Rizzo2013PRE}. The language of the overlap-functions in the replica symmetry breaking formalism has become one of the standards for explaining the physical nature of glasses forming events. For the Sherrington-Kirkpatrick (SK) model~\cite{sk1975} the glass transition problem in terms of RSB is explicitly solvable. Below we build the exact solution for the generalized SK-model without reflection symmetry using replica-symmetry breaking formalism.

One of the most interesting feature of spin-glass models is the connection between the replica method and dynamics~\cite{gotze2008complex,Rizzo2013PRE,Parisi2013JChemPhys,Parisi2010RevModPhys,Montanari2003,rizzo2013replica,KurchanJChemPhys2013}. The results we obtain below are in line with the results of Ref.~\cite{rizzo2013replica} when the 1RSB branch can be continued to FRSB branch of solutions of the Ising $p$-spin model.
In this connection it is worth mentioning about generalizations of SK model that include multispin (more than two) interactions. Then it was shown that the violation of FRSB scheme in generalized SK models is also correlated with the symmetry properties of the Hamiltonian~\cite{GoldbartSherringtonJPhys1985,NobreSherringtonJPhysA1993,Parisi2003,Marinari1998PRL,Ghirlanda1998JPhysA,guerra2003,Crisanti2004PRL}. There is a conjecture that in the absence of the reflection symmetry it is not possible to construct a continuous non-decreasing glass order parameter function $q(x)$ and so, the FRSB solution does not occur instantly in the point of RS solution instability unlike that is the SK model, see, e.g, Refs.~\onlinecite{Gross1985PRL,GribovaPRE2003}, where the Potts model is considered.

\section{Glass forming GSK-model\label{secmodel}}
Our Hamiltonian is a straightforward generalization of SK model~\cite{sk1975}: $\hat H=-\frac{1}{2}\sum_{i\neq j}J_{ij} \hat{U}_i \hat{U}_j$,
where  $i,j$ label the lattice cites. The exchange interactions have Gaussian distribution, $P(J_{ij})=\frac{\sqrt{N}}{\sqrt{2\pi} J}\exp\left[-(J_{ij})^{2}N/( 2J^{2})\right]$, where $J=\tilde{J}/\sqrt{N}$  and $N$ is the number of lattice sites. Using the replica-trick we can write down in standard way the disorder averaged free energy, order parameters and Almeida-Thoulles replicon modes $\lambda _{n{\rm\scriptscriptstyle RSB}}$ that indicate the $n$-th step of RSB  while $\lambda _{n{\rm\scriptscriptstyle RSB}} = 0$, see, e.g., Letter~\cite{Tareyeva2013PLA} and Refs. therein.

The bifurcation condition $\lambda_{\rm(\RS)\, repl}=0$ defines the point $T_{c}$ where the RS-solution becomes unstable. Considering respectively 1RSB, 2RSB, 3RSB, ..., $n\mathrm{RSB}$, we find that the equation $\lambda_{n{\rm\scriptscriptstyle RSB}} = 0$  always has the solution which determines the point $T_{c}$ and coincides with the solution of the equation, $\lambda_{\rm(\RS)\, repl}=0$~\cite{SchelkachevaPRB2010,TareyevaTMF2009,SchelkachevaJPhysA2011}.  We emphasize that it is the non-zero RS-solution that bifurcates.

To write $\Delta F$, the difference between the free energy and its replica symmetric value, as a  functional of Parisi FRSB glass order parameter, $q (x)$, and so to construct FRSB we use the standard formalized algebra rules~\cite{Parisi1980JPhysA5, mezard1987spin}. The properties of this algebra were formulated by Parisi for Ising spin glasses. In our case, the expansion of the generalized expression for the free energy includes some terms of non-standard form. Those terms are not formally described by the Parisi rules, but can be easily reduced to the standard form. To do this, we compare the corresponding expressions, consistently producing 1RSB, 2RSB, ... symmetry breaking. In what follows we use the complete equation for the free energy in this case up to fourth order in the deviations $\delta q ^{\alpha\beta}$ from $q_{\RS}$ where $ \alpha,\beta$ numbered replica (see Eq.(16) of Refs.~\cite{Tareyeva2013PLA}). Finally up to the terms of the third order in $q(x)$  we get near $T_{c}$ (fourth-order terms in $q(x)$ will be considered below):
\begin{widetext}
\begin{multline}\label{4440frs}
\frac{\Delta F}{Nk_BT}=-\frac{t^2}{4}\lambda_{\rm(\RS)\, repl} \langle q^{2}\rangle-\frac{t^{4}}{2}L\langle q\rangle^{2}-
\\
t^{6}\biggl[-B_{2}\langle q\rangle^{3} -B'_{2}\langle q\rangle^{3}+ B_{3}\left[\int_{0}^{1}zq^{3}(z)dz+3\int_{0}^{1}q(z)dz\int_{0}^{z}q^{2}(y)dy\right]+
B'_{3}\left[\langle q\rangle{\langle q^{2}\rangle}\right]+B_{4}\left[-{\langle q^{3}\rangle}\right]\biggr]+...
\end{multline}
where $t=\tilde{J}/kT$, $\langle q^{n}\rangle =\int_{0}^{1}q^n(z)dz$, for $n=0,1,\ldots$, FRSB glass order parameter $q_{\FRSB}= q_{\RS}+q(x) $. Coefficients in the Eq.~(\ref{4440frs}) are the averages of linear combinations of the products of operators $\hat U$ averaged on RS solution at the point $T_{c}$. Explicit total expression for $\Delta F$ we do not repeat here because it is rather lengthy, one can find it in Refs.~\cite{SchelkachevaJPhysA2011,Tareyeva2013PLA}. In particular
 \begin{align}\label{AARS}
 L=&\langle\hat{U}_{1}^2\hat{U}_{2}\hat{U}_{3}\rangle_{\RS}-
 \langle\hat{U}_{1}\hat{U}_{2}\hat{U}_{3}\hat{U}_{4}\rangle_{\RS}
\geq 0
\end{align}
where averaging is performed on RS solution at transition point. It follows from the Cauchy--Schwarz inequality that the expression Eq.~(\ref{AARS}) is nonnegative. The expression for $L$ is not equal to zero only when $q_{\RS}\neq 0$.

 The equation for the order parameter follows from the  stationarity condition $\frac\delta{\delta q(x)}\Delta F = 0$ applied to the free energy functional Eq.~(\ref{4440frs}). Since $\lambda_{\rm(\RS)\, repl}=0$ at $T_{c}$ we obtain:
\begin{multline}\label{1340frs}
-\frac{t_{c}^2}{2}\frac{d\left[\lambda_{\rm(\RS)\, repl}\right]}{dt}|_{t_{c}}\Delta t
q(x)-
t_{c}^{4}L\langle q\rangle-
\\
t_{c}^{6}\biggl[3\left(-B_{2}-B'_{2}\right)\langle q\rangle^{2}+3B_{3}\left(xq^{2}(x)+2q(x)\int_{x}^{1}q(z)dz+
\int_{0}^{x}q^{2}(z)dz\right)+
B'_{3}\left(\langle q^{2}\rangle+2q(x)\langle q\rangle\right)
-3B_{4}q^{2}(x)\biggr]+...=0.
\end{multline}

The non-trivial solutions of Eq.~(\ref{1340frs}) is  fulfilled only if
\begin{gather}\label{ten9788}
\langle q\rangle=0+o(\Delta t)^{2}.
\end{gather}
This is in fact the branching condition. It follows due to $L\neq 0$ in the expression Eq.~(\ref{1340frs}). As it can be seen from Eq.~(\ref{AARS}) this is a direct consequence of the fact that $q_{\RS}\neq 0$ if operators $\hat U$ do not possess the reflection symmetry. On the other hand, we know that 1RSB solution near the branch point satisfies the branching equation $\langle q\rangle_{\RSB}=0$, see  Refs.~\onlinecite{tareyeva2010spin,TareyevaTMF2009,SchelkachevaJPhysA2011}. This branching condition fails for $\hat{U}$ with reflection symmetry. For 2RSB solution we receive similar expression, $\langle q\rangle_{\mathrm{\scriptscriptstyle 2RSB}}=0$. Within our expansion in $\Delta t$ the results for 1RSB and 2RSB near $T_{c}$ coincide.  This leads eventually to Eq.~(\ref{ten9788}) for FRSB.

After the substitution in Eq.~(\ref{1340frs}), $x=0$ and $x=1$, and taking into account Eq.~(\ref{ten9788}), we obtain the following equation which will be needed later:
\begin{gather}\label{1319frs}
-\frac{t_{c}^2}{2}\frac{d\left[\lambda_{\rm(\RS)\, repl}\right]}{dt}|_{t_{c}}\Delta t
\left[ q(0)- q(1)\right]+
t_{c}^{6}3\biggl[B_{4}\left[q^{2}(0)-q^{2}(1)\right]+
B_{3}\left[q^{2}(1)+\langle q^{2}\rangle\right]
\biggr]+...=0.
\end{gather}\end{widetext}

Eq.~(\ref{1340frs}) can be further simplified
using the differential operator $\hat{O}=\frac{1}{q'}\frac{d}{dx}\frac{1}{q'}\frac{d}{dx}$, where $q'= \frac{d q(x)}{dx}$. Then $t^{6}\left\{B_{4}-B_{3}x\right\}+...=0$.

Now we get the key result --- the significantly depending on $x$ part of the solution $q(x)$ is concentrated in the neighborhood of
\begin{gather}\label{ten9888}
\tilde{x}={{B_{4}}/{B_{3}}}.
\end{gather}
We should repeat again that only for $\hat{U}$ with $\Tr{\hat{U}}^{(2k+1)}\neq0$, $k>0$ we get: $\tilde{x}={{B_{4}}/{B_{3}}}$. For operators with $\Tr{\hat{U}}^{(2k+1)}=0$ we get,  $B_{4}=0$ and $\tilde{x}=0$ which corresponds to the well-known result for SK and similar models with reflection symmetry.

Keeping the result~\eqref{ten9888} in mind we can now build FRSB. To describe the FRSB function $q(x)$  of the variable $x$ we have to include in the consideration the fourth-order terms in the expansion of $\Delta F$. This is in general done in Ref.~\cite{Tareyeva2013PLA}. Here we must reproduce this result to explain the origin of the FRSB solution that we build below:
\begin{widetext}
\begin{gather}\label{ten7688}
t^{6}\left\{B_{4}-B_{3}x\right\}+t^{8}\Biggl\{\left[-2D_{33}+4xD_{47}\right]\biggl[-xq(x)-\int_{x}^{1}dy q(y) \biggr]+
\left[-4D_{2}+2xD_{33}\right]q(x)+\left[D_{31}-D_{46}x\right]\langle q \rangle \Biggr\}=0\, ,
\end{gather}
\end{widetext}
where $\langle q \rangle=\int_{0}^{1}dy q(y)$ and  for the $D$-coefficients see  Ref.~\cite{Tareyeva2013PLA}.

The resulting equation (\ref{ten7688}) can be solved in a standard way as follows [we solve this equation formally  in the similar way as in Ref.~\cite{crisanti2010sherrington} where the expansion of the SK replica free energy functional around the nonzero RS solution, truncated to the fourth order, leads to a FRSB solution with continuous nonlinear order parameter]: We divide the equation by $\left[-2D_{33}+4xD_{47}\right]$, differentiate with respect to $x$, take into account Eq.~(\ref{ten9788}) and finally get:
\begin{gather}\label{ten955}
q(x)=\Gamma\frac{(x-s)}{\sqrt{(x-s)^{2}+\Delta}}-a,
\end{gather}
where
\begin{gather}\label{ten956}
s=\frac{D_{33}}{2D_{47}}, \\\Delta = -s^{2}+\frac{D_{2}}{D_{47}}, \\ a=\frac{1}{2t_{c}^{2}}\frac{(B_{3}D_{33}-2B_{4}D_{47})}
{(-D^{2}_{33}+4D_{2}D_{47})},
\end{gather}
and $x$ is near $\tilde{x}$. The values of $\Gamma$ and $x_{c}$ (the boundary value of $x$) should be found from the initial conditions.

It should be noted that our conclusions are consistent with Ref.~\onlinecite{rizzo2013replica} where static and dynamics of a class of mean-field spin-glass models were considered. It was shown earlier that it may exist a temperature at which the stable at higher temperatures 1RSB branch becomes unstable at lower temperatures and it  can be continued to a FRSB branch. An analytical study of the fourth order expansion of the free energy in the context of some truncated model reveals that the FRSB branch of solutions is characterized by the two plateau structure and the continuous region. Numerical solution of the FRSB equations for the Ising $p$-spin model  with $p = 3$ have been obtained where $ q(x)$ depends on $x$ in a nonlinear manner. This is essentially a generalization  of the result obtained originally by Kanter, Gross and Sompolinsky in the context of the Potts  model~\cite{Gross1985PRL}.

Next, we proceed by successive steps. From Eq.~(\ref{ten9888}) follows that $\tilde{x}\equiv \tilde{x}_{\1RSB}$: i.e. the value at which the 1RSB solution changes abruptly from $q_{\1RSB}(0)$ to $q_{\1RSB}(1)$, see  Refs.~\onlinecite{tareyeva2010spin,TareyevaTMF2009,SchelkachevaJPhysA2011}. We obtained that within our expansion in $\Delta t$ the results for 1RSB and 2RSB near $T_{c}$ coincide. Therefore, we start from the 1RSB solution, which is already a good approximation~\cite{mezard1987spin,JanisPRB2011}. We recall here that 1RSB solution behaves much differently when operators $\hat{U}$ have reflection symmetry, and when they do not have such symmetry (strict detailed derivation is given in ~\cite{SchelkachevaJPhysA2011} for $p=2$). At first we obtain from Eqs.~(\ref{ten9788})-(\ref{ten9888}) the values of $\tilde{x}$, $q_{\1RSB}(0)$ and $q_{\1RSB}(1)$. General equations ~(\ref{ten9788})-(\ref{ten9888}) can easily be rewritten, assuming any nRSB wherein $\tilde{x}= \tilde{x}_{\mathrm{\scriptscriptstyle nRSB}}$. Further we use  Eq.~(\ref{ten9788}) as
\begin{gather}\label{ten9555}
q_{\1RSB}(0)\tilde{x}+q_{\1RSB}(1)(1-x_{c})+\int_{\tilde{x}}^{x_{c}}dy q(y)=0,
\end{gather}
and equation Eq.~(\ref{ten7688}) for $x=\tilde{x}$:
\begin{multline}\label{ten7685}
\left[-D_{33}+2\tilde{x}D_{47}\right]\left[-\tilde{x}q(\tilde{x})+q_{\1RSB}(0)\tilde{x} \right]+
\\
\left[-2D_{2}+\tilde{x}D_{33}\right]q(\tilde{x})=0.
\end{multline}
Finally from the equations (\ref{ten9555})-(\ref{ten7685}) we find $\Gamma$ and  $x_{c}$.

\begin{figure}[t]
  \includegraphics[width=0.9\columnwidth]{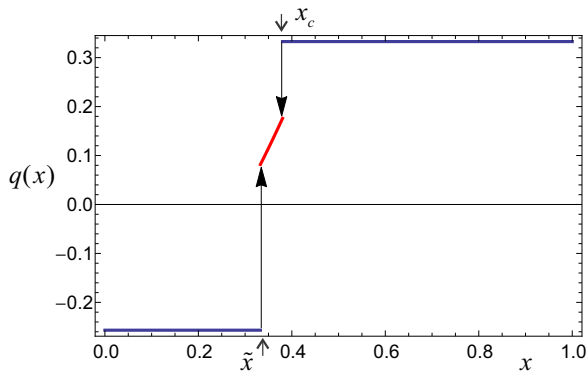}\\
  \caption{(Color online) Order parameter $q(x)$ is defined by the expression (8) where $\hat U = \hat S +
0.5 \hat Q$ and $(T-T_{c})=-0.3$. ($\hat S$ is the $z$-component of spin  (for ${\bf S}=1$), $\hat Q$ is the axial quadrupole moment $\hat Q = 3 {\hat S}^2 -2$.) Horizontal sections are, respectively, $q_{\1RSB}(0)$ and $q_{\1RSB}(1)$. Function $q_{\FRSB} = q_{\RS} + q(x)$, where $q_{\RS}|_{T=T_{c}}=1.154.$}\label{figq0_q1_VQ}
\end{figure}

\section{Discussion}

As an example, we consider operators $\hat U = \hat S + \eta \hat Q$ where $\eta$ is a tuning parameter, not small. Here $\hat S$ is the $z$-component of spin  (for ${\bf S}=1$) taking values $(0,1,-1)$, $\hat Q$ is the axial quadrupole moment, $\hat Q = 3 {\hat S}^2 -2$, and it takes values $(-2, 1, 1)$ (see,~e.g.~\cite{schelkachevaPLA2006}). Algebra of the operators {$\hat Q, \hat S, E$} is closed: ${\hat Q}^2 = 2 - \hat Q$, $3 {\hat S}^2= 2 + \hat Q$, and $\hat Q \hat S = \hat S \hat Q = \hat S$. The operator $\hat S$ has the reflection symmetry while $\hat Q$ has not. FRSB is valid for reflection symmetric operator $\hat S$ ~\cite{schelkachevaPLA2006}. The operator $\sqrt{3}S = V$ is a second component of the quadrupole momentum operator considered in the problem of anisotropic quadrupole glass.

For $\eta=0.5$ (see Fig.2) we obtain $T_{c}=1.809$; $q_{\RS}|_{T_{c}}=1.154$; $\tilde{x}=0.333$, $x_{c}=0.392$ for $(T-T_{c})=-0.3$. For $(T-T_{c})=-0.1$ we have $x_{c}=0.35$. For $(T-T_{c})=-0.2$ we obtain $x_{c}=0.37$. In the case of $\hat U =\hat Q$ we obtain $T_{c}=1.37$ and $\tilde{x}=0.43$, $x_{c}=0.446$ for $(T-T_{c})=-0.2$. Since $q_{\RS}\neq 0$ we have for FRSB glass order parameter $q_{\FRSB}= q_{\RS}+q(x) $. We remind here that in the case of the standard truncated SK-model, well-known Parisi  FRSB solution is $q_{\FRSB}= q(x) = x/2$ for $0 \leq x\leq (-2)(T-T_{c})$ and $q_{\FRSB}= -(T-T_{c})$ for $x > (-2)(T-T_{c})$ where $T_{c}=1$, see, e.g.~Ref.~\cite{mezard1987spin}.

So, in terms of qualitative physical arguments we can define in conventional manner the distribution function $P(q)$  as the order parameter, which gives the probability of finding a pair of glass states  having the overlap $q$. In terms of the FRSB scheme the distribution function $P(q)$ is defined by the Parisi  function $q(x)$: $P(q) = dx(q)/dq$.  So the continuous spectrum of the overlaps appears in the whole interval  ${\tilde x(T)} < x < x_{c}(T)$. When non-symmetric part of $\hat U$ is small our solution for $q(x)$ is linear~\cite{Tareyeva2013PLA}) like SK-model solution in the presence of a small external field~\cite{mezard1987spin}.

The proximity of the boundary parameter $x_{c}$ to $\tilde{x}$  says that 1RSB solution is generally quite good physical approximation for the problem we are considering. In this regard, our solution is close to that obtained in Ref.~\onlinecite{Crisanti2004PRL} and in a series of subsequent papers where the phase diagram was presented of the spherical $2 + p$ spin glass model. The main outcome is the presence of a new phase with an order parameter made of a continuous part much alike the FRSB order parameter and a discontinuous jump resembling the 1RSB case.

 To say for sure whether the presence of such a jump is an  intrinsic feature of our model or we need to use next subsequent successive steps, it is necessary to consider the further approximation for free energy up to next order in $(\Delta T)$.

\section{Conclusions \label{Sec:Conc}}
To summarize, we have considered a model  with pair interaction  where the absence of reflection symmetry is caused by the characteristics of the operators $\hat{U}$ themselves.  The principal prescription for obtaining a full replica symmetry breaking solution is derived in general case in the neighbourhood of the transition temperature. An illustrating example is considered demonstrating the explicit build solution.

\acknowledgments

This work was supported in part by the Russian Foundation for Basic Research (Grants No. 11-02-00341 and 13-02-91177), the Russian Academy of Sciences programs, the Grant of President of Russian Federation for support of Leading Scientific Schools No.~6170.2012.2 and NSF Grant DMR 1158666.

\bibliography{references}
\end{document}